# Mining and Analyzing Patron's Book-Loan Data and University Data to Understand Library Use Patterns

**Tipawan Silwattananusarn**
Department of Library and Information Science
Faculty of Humanities and Social Sciences
Prince of Songkla University, Pattani Campus
Pattani, Thailand
Corresponding Author: tipawan.s@psu.ac.th

**Pachisa Kulkanjanapiban**
Khunying Long Athakravisunthorn
Learning Resources Center, Prince of
Songkla University, Hatyai, Songkla,
Thailand

## Abstract

The purpose of this paper is to study the patron's usage behavior in an academic library. This study investigates on pattern of patron's books borrowing in Khunying Long Athakravisunthorn Learning Resources Center, Prince of Songkla University that influence patron's academic achievement during on academic year 2015-2018. The study collected and analyzed data from the libraries, registrar, and human resources. The students' performance data was obtained from PSU Student Information System and the rest from ALIST library information system. WEKA was used as the data mining tool employing data mining techniques of association rules and clustering. All data sets were mined and analyzed to identify characteristics of the patron's book borrowing, to discover the association rules of patron's interest, and to analyze the relationships between academic library use and undergraduate students' achievement. The results reveal patterns of patron's book loan behavior, patterns of book usage, patterns of interest rules with respect to patron's interest in book borrowing, and patterns of relationships between patron's borrowing and their grade. The ability to clearly identify and describe library patron's behavior pattern can help library in managing resources and services more effectively. This study provides a sample model as guideline or campus partnerships and for future collaborations that will take advantage of the academic library information and data mining to improve library management and library services.

## Introduction

In the academic library, with the continuous advancement of the automation and digitization, the amount of data continues to be collected and stored in centralized or distributed databases. Some academic libraries are using various information technologies to manage the massive patron's data such as borrowing information of books in the library. The library patrons are the core targets of the services in the academic library (Siguenza-Guzman, Saquicela, Avila-Ordonex, Vandewalle & Cattrysse, 2015). Prediction of the service demand and comprehensive analysis of patron's behavior are truly of great value to the construction of the resources of academic library and the construction of disciplines in universities (ibid).

Most academic libraries are unable to find the connection and pattern of various types of data in their database, unable to predict future trends of the services based on existing data, and lack ability of data mining to mine the hidden knowledge behind their own data (Chen, 2014). Data mining techniques when applied to library information repositories helps to discover interesting patterns of data that help librarians to take decision on books selection(Krishnamurthy & Balasubramani, 2014), customer segmentation or identification of patterns of customer behavior, providing the appropriate library services and collections, and responding the needs of patrons(Krishnamurthy & Balasubramani, 2014; Siguenza-Guzmanet al., 2015).

Khunying Long Athakravisunthorn Learning Resources Center is an academic library affiliated to the Prince of Songkla Universityon Hat Yai campus (Facilities on campus, Prince of Songkla University, 2020). The library has implemented an in-house library automation system, the Automated Library System for Thai Higher Education Institutes (ALIST), since2004. As technology continues to evolve, the library has entered the information age. While the primary mission of the library has been supporting the teaching and research, the library has been urged to redesign its programs and services. A trend for the computerized level in the library has continued to increase. Data analytics has become a critical component of most library research. Data mining techniques propose to analyze a large data set for finding insights patterns of library activities and user achievement. This makes it possible to predict the future academic library performance based on these sources of information. The library has been collecting data and used these data to answer questions, for example, the number of borrowing, the popularity of various collections, budget allocation for acquisition, etc. to accomplish the objectives ensuring the availability in the library. There will be great potential in the data collected by the library from using data mining methods to produce services for users, assist with decision making, and provide library patterns impact on academic achievement and research.

The main aim of this study is to identify the patterns of how university students, staff, and personnel utilize library services and resources. Data mining techniques – association rules and clustering analysis – have been proposed to analyze the data collected by the automated library system, registration, and human resources. Therefore, how to analyze the needs of the library patrons more effectively is especially important to provide patrons with more personalized needs and services, and to provide librarians with more reliable decision supports.

## Review of Literature

Several research works used library analytics to extract interesting facts such as library patron behavior. These works were possible due to the effective use of the information technology to ease the data management and information retrieval. Most of them have common approach and objective, i.e. applying data mining techniques to understand and track the behavior of individuals or certain groups. Study of data mining from the databases of the library system would be useful in helping predict library patron's usage patterns and infer decision through the knowledge gained. Findings from Peking University library (Yan, Zhang, Tang,

Sun, Deng & Xiao, 2010) showed that mining users' book-loan data could build links of a book-borrowing between users and books and could build links of a book-borrowing between schools and degrees. The users' needs at library of Changchun Institute of Technology (Zhu & Zhang, 2011) can be analyzed effectively with data mining decision tree C4.5 algorithm. The results of mining showed some potential links among the factors affecting the using of library digital resources. Processing the circulation data of North South University Library (Hossain & Rahman, 2015) using statistics and association rules has made the analysis easier for calculating library material acquisition and efficient management of budget allocation.

The development of the predictive analytic framework has been used by academic libraries to gain a valuable insight. The classification results of data mining applications in academic libraries were presented and discussed in Siguenza-Guzman et al. (2015). Chen (2014) designed a set of model structure combined with university libraries system in order to analyze the related knowledge of personalized information system based on data mining technology. In order to understand students' needs, association rules mining was applied to library circulation data to find interesting rules with respect to users' demands in usage of books (Krishnamurthy & Balasubramani, 2014; Lipeng & Wang, 2015).The analysis of library and university data using descriptive statistics analysis and correlation analysis in Renaud, Britton, Wang & Ogihara (2015) revealed patterns of library use by academic department, patterns of book usage over 20 years and correlations between library use and grade point average. Findings from the use of data mining from library management system by *K*-means algorithm of the cluster analysis (Wang, Tang, Liu & Li, 2011; Hajek & Stejskal, 2014) showed that the cluster results satisfied the demand of the student readers and guided the library to adjust service strategy and provide personalized services. In view of characteristics of users' data in the university library and based on big data technology, Cheng & Liu (2019) suggested that big data mining for collection resources and reader information requires more staffs with in-depth professional skills of data mining and should have a good sense of innovation and data mining inspiration.

The application of data mining technology to the various services of academic libraries has gradually become the mainstream trend (Huancheng et al., 2018). In recent years, Cao et al. (2018) stated that data mining is one of the cores of the current development of smart library technology. Impacts of data mining on academic libraries have been included search and resource discovery, scholarly publishing, learning, acquisition and curation, infrastructure building, user navigation, and data literacy (Cox et al., 2019). Rattan (2019) identified the core library areas where data mining techniques can be applied to build a stronger serviceable library system for the maximum benefit of library users. Libraries through data mining techniques would be able to strengthen its managerial and decision support system (Rattan, 2019). Library book usage has shown a significant contribution to academic success and student retention. Ochilbek (2019) stated that library usage and book borrowing are important factors in a student's academic performance. Shi & Zhu (2015) use data detection to identify academic library use patterns and to judge whether students' GPA correlated to academic library use. Academic performance is not only predictable from the book-loan history but also improves the recommendation of library books for students (Lian, Ye, Zhu, Liu, Xie, & Xiong, 2016).

Our work differs from related works studied by applying data mining tools based on the feature of *Apriori* association rules algorithm and *K*-Means clustering algorithm to explore explanatory knowledge and derive appropriately utilization gain.

## Methods and Materials

Here, describes the methodology used for analyzing circulation data using association rules mining technique to identify patron's interest, and analyzing the characteristics of the collections of different user groups and analyzing the characteristics of each user to obtain the valuable information and individual needs using clustering technique. The flow of methodology is shown in Figure 1. The methodology shownconsists of 3 important stages, i.e. data collection, data preprocessing, and data mining modeling.

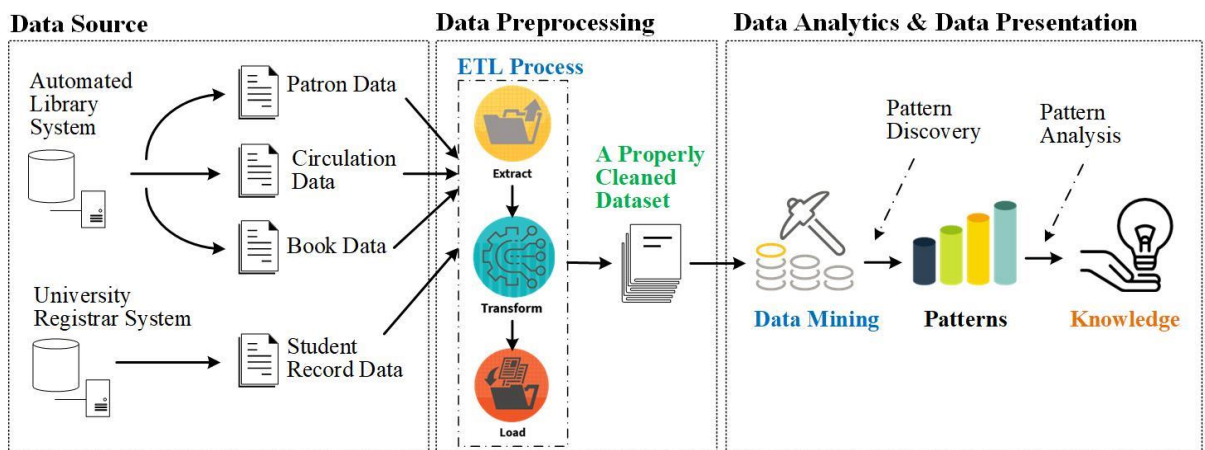

Figure 1: *Academic Library Analytics Framework for Patron's Book-Loan Behavior Prediction.*

## Data collection

The data collected from the data resource platforms of (1) Library Information System (ALIST), Khunying Long Athakravisunthorn Learning Resources Center, Prince of Songkla University (PSU) and (2) PSU Student Information System.

The data collected includes the following data sets:

### Patron data

This dataset includes all university student and staff information (43,550 records). Data fields includepatron barcode, patron type, faculty ID, total number of book checkouts, and other related information.

### Book data

This dataset contains about 168,814 records of item data of book collection in the ALIST system. Important data fields are item barcode, title, author, catalog date (created date), last check-in date, last check-out date, total times check-out.

### Circulation data

This dataset contains loan information about books checked out during academic years2015-2018. There is a total of 102,929 transactions. The data fields include patron barcode and item barcode for currently check-out items during an academic year.

### Student record data

This dataset contains data of students'academic achievement in the Student Information System. There are 64,064 registered students during academic years 2015-2018. Data fields includestudent ID, student name, faculty ID, cumulativeGPA (CGPA).

**Data preprocessing**

    Data preprocessing is an initial step to process the collected raw data and prepare for data mining. The steps include data cleansing, data extraction, and data integration. The datasets from ALIST library database were modified for our research as given in Table 1:

Table 1
*Data sets attributes*

| Original datasets | Modified datasets for our research |
|---|---|
| 1.  Book data | Items (item_barcode, category_code, catalog_date) |
| 2.  Patron data | Patrons (patron_barcode, faculty_id) |
| 3.  Circulation data | Circulation (item_barcode, patron_barcode, lifespan) |
| 4.  Student record data | Student (student_ID, faculty_id, CGPA) |
| 5.  Category data[*] | Category (category_code, description) |

[*] *There is no dataset for category. It was created based on LC classification.*

    Since there is no category data in ALIST, the Category dataset was created based on Library of Congress Classification (LC Classification). In circulation data, attribute of lifespan can be expressed as the length of time from the date of cataloging to the date of last check-in. There are four groups of lifespan data: 0-5 years, 6-10 years, 11-15 years, and 16-20 years. For student record data, values of cumulative grade point average (CGPA) are grouped into five levels of student academic achievement status (Excellent, Very Good, Good, Average, Poor) – Excellent: with CGPA range of 3.50-4.00, Very Good: with CGPA range of 3.00-3.49, Good: with CGPA range of 2.50-2.99, Average: with CGPA range of 2.00-2.49, and Poor: with CGPA less than 2.00.

**Data Mining Model - Mining Patron's Book-Loan Data using Association Rules and Clustering Approaches**

    In this study, the main models for data mining include association analysis and cluster analysis. Association rules are currently the most widely used method (Chen, 2014; Witten et al., 2017).The principle of association rules is to discover the correlation between two things or many things. Association rules mining of books are to look for potential borrowing rules from patrons. For example, by association analysis, it can be found that 70% of patrons who borrow the book *"Flora of Thailand"* also borrow the *"Taxonomy of flowering plants"*. Cluster analysis can divide the records in dataset into a series of meaningful groups, and then perform statistics and description work on them. In data mining, clustering methods is used to identify certain groups of objects with similar characteristics. The *K*-Means algorithm is used for the identification of services of the public library (Hajek & Stejskal, 2014). For example, patrons in cluster $c_1$ is the least demanding on the library services; while patrons in cluster $c_2$ also use other services offered by the library.

***Association Analysis Model with Apriori***

    Association rules are used to predict correlation between the data items (Krishnamurthy & Balasubramani, 2014).Rules can predict any attribute or combination of attributes. *Apriori*is the

standard association rules algorithm and minimum confidence value must be specified to seek rules with the most support (Witten et al., 2017).Support refers to the number of instances that appears in the rule; while confidence refers to proportion of instances that appears at the left-hand side for which the right-hand side also holds (ibid). The key concept of *Apriori* algorithm is to search through a subset of a frequent item set and uses frequent items to generate association rules (ibid). The strategy of *Apriori* algorithm is iteratively reduce the minimum support until the required number of rules is found with a given minimum confidence (ibid).

### *Cluster Analysis Model with K-Means*

Cluster analysis is defined as grouping a set of similar objects into clusters (Bhatia, 2019).The core idea of clustering is trying to divide the instances into clusters so that instances within intra-cluster have high similarity with one another but have high dissimilarities in inter-cluster (Witten et al., 2017;Yu, 2011; Bhatia, 2019). In this study, *K*-Means clustering algorithm is used in analyzing patron's characteristics. In *K*-Means clustering algorithm, the objects are grouped into *k* number of clusters based on attributes or features. The grouping of objects is done by minimizing the sum of squares of distances between data and the corresponding cluster centroid (Bhatia, 2019). The working of *K*-Means algorithm is stated as follows: specify the desired number of clusters *k*, choose *k* points at random as cluster centers, assign all instances to their closest cluster center, calculate the centroid of instances in each cluster, these centroids are the new cluster centers, and continue until the cluster centers no more changes (Witten et al., 2017; Bhatia, 2019).

## Results and Discussion

The most interesting part of this study is to observe the patterns of library users. Three major issues were investigated. First, *who uses the library most frequently? Do patrons from different faculties borrow different category of books? If their interests are different, how category is related to faculty?* Second, *what is the useful lifespan of a book? Which category has long or short life span?* Finally, *what are the association rules related to patrons borrowing books? Is there relationship between student academic achievement and their book loans?* Using the data collected, descriptive statistical and data mining methods were applied to analyze for patron's behavior and patron's interest, the association patterns of library category usage and the relationship between patron's cumulative grade point average and their library use behavior.

Analysisof patron behavior and patron interest

To discover who the most frequent user of the library is, the analysis of book loan patterns by patron type was carried out. The percentage of check-out distributed by patron type during academic years 2015-2018 is presented in Table 2. The percentage of check-out is determined from the number of patron type's check-out and the total number of check-out books. It is evident from Table 2 that undergraduates have the highest percentage of check-out(shown in bold with 63.14%, 72.64%, 69.23%, 69.46% respectively) followed by graduate students, others and then the academic staff. This indicates that the percentage of check-outs roughly corresponds to the size of patron types, i.e. undergraduates is the group with the highest checking out items, followed by graduate students and then others. Table 3 also shows that book collection is the most commonly check-out material (shown in bold). The subsequent collection type is quite different depending on the patron type, i.e. Facilities and equipment (shown in purple) for undergraduates, while theses (shown in blue) takes the second place for graduate students, and fictions (shown in green) for the academic staff. Table 4 shows the difference in check-out collections for undergraduate students. For books, the sophomore students have the higher (65.26%) checking outs than freshmen (61.56%), junior (62.75%), and senior students (60.49%). This pattern is not hold for facilities and equipment which are much more used among undergraduates. The first-year students use items in facilities and equipment collection (25.46%) more than second year (23.96%), third year (32.49%), and senior students (17.27%).Junior students are the top in borrowing fiction, while senior students'top reading lies in theses collection. Table 5 shows the top ten faculties ranked by percentage of total book check-outs. Top ten faculties list in Table 5 are based on the proportion of number of book check-out by undergraduate students in the faculty to the total number of book check-out. Faculty of Science (shown in blue) and Faculty of Engineering (shown in green) take the top-level and the second-level ranking with a large gap. It should be noted that Faculty of Medicine, Faculty of Pharmaceutical Sciences, and Faculty of Nursing are also in the top ten, though these faculties have their own libraries and book loans records in their system not included in this study.

Table 2
*Total check-out distributed by patron type during academic year*

| Patron types | % of total check-outs(number of patron types) | | | |
|---|---|---|---|---|
| | Academic Year 2015 (1/8/2015 – 31/7/2016) | Academic Year 2016 (1/8/2016 – 31/7/2017) | Academic Year 2017 (1/8/2017 – 31/7/2018) | Academic Year 2018 (1/8/2018 – 31/7/2019) |
| **Undergraduate** | **63.14 (16,379)** | **72.64 (16,478)** | **69.23 (16,158)** | **69.46 (15,846)** |
| Graduate | 18.95 (3,690) | 13.30 (3,050) | 15.54 (2,685) | 13.72 (2,486) |
| Academic Staff | 5.24 (1,523) | 4.89 (1,552) | 5.03 (1,591) | 6.33 (1,581) |
| Others | 12.67 (7,312) | 9.17 (7,331) | 10.20 (7,323) | 10.49 (7,431) |
| Total | 100.00 | 100.00 | 100.00 | 100.00 |

Table 3
*Check-out distribution by collection and patron type during academic year*

| Collection | % of total check-outs | | | | | | | | | | | |
|---|---|---|---|---|---|---|---|---|---|---|---|---|
| | Academic Year 2015 | | | Academic Year 2016 | | | Academic Year 2017 | | | Academic Year 2018 | | |
| | UG | G | AS | UG | G | AS | UG | G | AS | UG | G | AS |
| **Books** | **79.45** | **70.37** | **73.95** | **55.97** | **72.72** | **67.29** | **57.98** | **71.96** | **74.32** | **59.62** | **67.43** | **70.43** |
| Multimedia | 1.21 | 3.29 | 3.66 | 0.70 | 4.36 | 7.57 | 6.17 | 2.66 | 5.34 | 7.69 | 2.74 | 3.38 |
| Serials | 1.87 | 4.51 | 4.01 | 1.03 | 3.19 | 6.02 | 1.02 | 4.60 | 3.85 | 0.73 | 2.95 | 4.21 |
| Projects | 1.43 | 0.83 | 0.49 | 1.04 | 0.48 | 0.41 | 0.75 | 0.46 | 0.16 | 0.56 | 0.72 | 0.15 |
| **Theses** | 2.54 | **14.68** | 1.13 | 2.01 | **13.30** | 2.97 | 1.73 | **15.62** | 0.84 | 1.28 | **19.33** | 1.20 |
| **Fictions** | 6.91 | 3.97 | **11.88** | 6.04 | 2.40 | **11.26** | 7.23 | 2.37 | **10.35** | 6.63 | 2.92 | **10.64** |
| Juvenile | 0.16 | 1.33 | 4.60 | 0.25 | 0.18 | 3.52 | 0.20 | 0.18 | 4.72 | 0.20 | 0.33 | 9.47 |
| **Facilities & Equipment** | **6.15** | 0.52 | 0.23 | **32.84** | 3.05 | 0.96 | **24.86** | 1.97 | 0.42 | **23.23** | 3.50 | 0.49 |
| Others | 0.27 | 0.50 | 0.06 | 0.12 | 0.32 | 0.00 | 0.06 | 0.19 | 0.00 | 0.07 | 0.09 | 0.03 |

NoteUG = Undergraduate, G = Graduate, AS=Academic Staff; Books = Book in Thai, Textbook, Reserved book, SET Corner, Information Southern Thailand-IST, Asean, Moral, References, PSU Publication; Multimedia = Video CD, DVD, CD; Multimedia, Audio CD, Movie, Tape Cassette, Kit; Serials = Serials; Projects = Project; Theses = Thesis written in Thai, Thesis, Minor Thesis; Fictions = Fiction written in Thai, Fiction; Juvenile = Juvenile; Facilities & Equipment = iPad, IT Accessories; Others = British Library, Standard Pub., Pamphlet

Table 4

*Collection check-out distributed by undergraduate class during academic year 2015-2018*

| Undergraduate Class | Collections | % of total check-out | | | | |
|---|---|---|---|---|---|---|
| | | 2015 | 2016 | 2017 | 2018 | Total |
| Freshmen | | | | | | |
| | Books | 82.16 | 53.58 | 54.94 | 63.70 | 61.56 |
| | Multimedia | 0.56 | 0.27 | 10.66 | 12.68 | 5.72 |
| | Serials | 1.13 | 0.47 | 0.74 | 0.17 | 0.61 |
| | Projects | 0.24 | 0.07 | 0.10 | 0.02 | 0.10 |
| | Theses | 0.23 | 0.04 | 0.24 | 0.28 | 0.18 |
| | Fictions | 7.55 | 5.08 | 5.86 | 7.10 | 6.19 |
| | Juvenile | 0.13 | 0.04 | 0.12 | 0.16 | 0.10 |
| | F&E | 7.78 | 40.43 | 27.29 | 15.83 | 25.46 |
| | Others | 0.22 | 0.03 | 0.04 | 0.06 | 0.08 |
| | Total | 100.00 | 100.00 | 100.00 | 100.00 | 100.00 |
| Sophomore | | | | | | |
| | Books | 81.77 | 56.75 | 64.74 | 59.42 | 65.26 |
| | Multimedia | 0.68 | 0.22 | 5.05 | 8.84 | 3.28 |
| | Serials | 1.06 | 0.57 | 0.38 | 1.41 | 0.81 |
| | Projects | 0.20 | 0.43 | 0.07 | 0.12 | 0.23 |
| | Theses | 0.68 | 1.15 | 0.36 | 0.43 | 0.70 |
| | Fictions | 6.39 | 6.45 | 4.97 | 3.46 | 5.46 |
| | Juvenile | 0.08 | 0.49 | 0.09 | 0.05 | 0.20 |
| | F&E | 8.89 | 33.87 | 24.27 | 26.23 | 23.96 |
| | Others | 0.24 | 0.06 | 0.06 | 0.05 | 0.10 |
| | Total | 100.00 | 100.00 | 100.00 | 100.00 | 100.00 |
| Junior | | | | | | |
| | Books | 80.92 | 60.05 | 55.36 | 57.69 | 62.75 |
| | Multimedia | 1.07 | 0.77 | 3.69 | 4.64 | 2.48 |
| | Serials | 1.86 | 0.95 | 0.72 | 0.49 | 0.98 |
| | Projects | 1.21 | 0.68 | 0.26 | 0.35 | 0.61 |
| | Theses | 2.13 | 1.16 | 2.02 | 0.89 | 1.54 |
| | Fictions | 7.75 | 6.20 | 9.77 | 7.88 | 7.86 |
| | Juvenile | 0.08 | 0.42 | 0.10 | 0.20 | 0.21 |
| | F&E | 4.81 | 29.67 | 28.01 | 27.85 | 23.49 |
| | Others | 0.17 | 0.10 | 0.07 | 0.01 | 0.09 |
| | Total | 100.00 | 100.00 | 100.00 | 100.00 | 100.00 |
| Senior | | | | | | |
| | Books | 73.16 | 55.02 | 57.24 | 56.63 | 60.49 |
| | Multimedia | 2.54 | 1.85 | 2.92 | 3.22 | 2.59 |
| | Serials | 3.49 | 2.49 | 2.53 | 0.90 | 2.39 |

| Undergraduate Class Freshmen | Collections | % of total check-out | | | | |
|---|---|---|---|---|---|---|
| | | 2015 | 2016 | 2017 | 2018 | Total |
| | Projects | 4.13 | 3.59 | 3.03 | 1.82 | 3.19 |
| | Theses | 7.16 | 6.82 | 5.41 | 3.63 | 5.85 |
| | Fictions | 6.18 | 6.88 | 9.59 | 8.17 | 7.63 |
| | Juvenile | 0.34 | 0.14 | 0.57 | 0.41 | 0.35 |
| | F&E | 2.57 | 22.87 | 18.64 | 25.07 | 17.27 |
| | Others | 0.44 | 0.34 | 0.07 | 0.15 | 0.26 |
| | Total | 100.00 | 100.00 | 100.00 | 100.00 | 100.00 |

* F&E = Facilities and Equipment

Table 5
*Total check-out distributed by undergraduate's faculty*

| Academic Year 2015 | | Academic Year 2016 | | Academic Year 2017 | | Academic Year 2018 | |
|---|---|---|---|---|---|---|---|
| Top 10 Faculty | % of total check-out | Top 10 Faculty | % of total check-out | Top 10 Faculty | % of total check-out | Top 10 Faculty | % of total check-out |
| Science | 26.48 | Science | 28.18 | Science | 24.67 | Science | 28.00 |
| Engineering | 17.96 | Engineering | 18.83 | Engineering | 17.12 | Engineering | 17.40 |
| Natural Resources | 10.81 | Natural Resources | 10.60 | Natural Resources | 10.47 | Management Sciences | 10.20 |
| Management Sciences | 8.24 | Management Sciences | 7.16 | Management Sciences | 9.41 | Natural Resources | 9.48 |
| Liberal Arts | 6.67 | Liberal Arts | 5.49 | Pharmaceutical Sciences | 5.46 | Liberal Arts | 6.01 |
| Agro-Industry | 5.11 | Law | 4.85 | Liberal Arts | 5.38 | Law | 5.24 |
| Law | 4.54 | Pharmaceutical Sciences | 4.23 | Law | 5.26 | Pharmaceutical Sciences | 4.58 |
| Pharmaceutical Sciences | 4.28 | Economics | 4.18 | Agro-Industry | 4.17 | Nursing | 3.48 |
| Economics | 3.70 | Nursing | 3.62 | Medicine | 3.54 | Traditional Thai Medicine | 2.94 |
| Nursing | 3.44 | Agro-Industry | 3.57 | Nursing | 3.39 | Medicine | 2.92 |
| Others | 3.36 | Others | 5.34 | Others | 6.39 | Others | 4.11 |

## Analysis of patron's book loan behavior

Book loan patrons were grouped based on undergraduate status, faculty and LC category. There are patrons from 17 different faculties and 21 LC subject categories involved in the book loan data. The interested patterns of patrons in book-borrowing are: *"Does patrons in each faculty tend to borrow subject relevant category?"* and *"How categories draw patrons from subject irrelevant category?"*

First, how each faculty influences patrons in borrowing subject relevant category was investigated. Table 6 shows an overview of the percentage of borrowing relationship from 17 faculties to 21 LC books categories. The calculated percentage equals the number of patrons in a faculty who checked out divided by the total number of check-outs. In Table 6 faculties and categories of books are represented by IDs. The meaning of each ID corresponding to faculty and book category is shown in Table 7. It is clear from Table 6 that patrons in each faculty tend to borrow books relevant to subject category. For example, patrons from Faculty of Law (*ID:5*) borrow the highest number of books in category *K:Law*; while patrons from Faculty of Natural Resources (*ID:10*) borrow higher number of books in category *S:Agriculture*. It is also interesting to note that patrons from the health sciences groups, including the Faculty of Medicine, Nursing, Dentistry, Pharmaceutical Sciences, Veterinary Science, Traditional Thai Medicine, and Medical Technology, have significant check-outs of books in category *Q: Science* and category *R: Medicine*. However, Table 6 shows that patrons from Faculty of International College, Engineering, and Agro-industry tend to borrow books in category *T: Technology*. Curriculums in each faculty may be an important factor for this phenomenon. Faculty of International College, for example, consists of only one program in digital media; while Faculty of Agro-Industry has three programs, i.e. Food Science and Technology, Material and Packaging Technology, and Agro-Industry Technology and Management (International Affairs Office, Prince of Songkla University, 2019).

How subject irrelevant categories attract patrons was also investigated. The knowledge dependency between faculty and LC category was sort out by analyzing what influences in each category. Table 7 shows the faculty, LC Category, and faculty-LC dependency (column 3). Non-technology book categories borrowed by health sciences faculties, i.e. Traditional Thai Medicine, Medical Technology, Medicine, and Nursing are categories *B, D, G, L, N, and P*. It is also shown that students in Traditional Thai Medicine like to borrow books in non-technology categories most. However, students in Faculty of Liberal Arts and Management Sciences have some similar check-out patterns in terms of subject. This is similar to the results of other study. The top 5 popular non-tech book categories borrowed by Science and Engineering major students are categories *F: Economics*, *I: Literature*, *K: History and Geography*, *D: Politics and Law*, and *B: Philosophy and Religion*; while the top 5 popular non-humanity book categories borrowed by Liberal Arts and Humanities students are categories *TP: Computer Technology*, *R: Medical Science*, *O2: Statistics*, *O1: Mathematics*, and *TU: Architecture Science* (Yan el al., 2010).

It is interesting to realize that their check-out books fall in non-technology categories. In terms of widely influential patterns from patron's book borrowing, Table 7 presents that books in categories *B: Philosophy, Psychology, Religion, P: Language and Literature, D: World History, and A: General Works* are widely borrowed.

Table 6
*Percentage of patron loan by faculty and LC class*

| IDs | 1 | 2 | 3 | 4 | 5 | 6 | 7 | 8 | 9 | 10 | 11 | 12 | 13 | 14 | 15 | 16 | 17 |
|-----|------|------|-------|------|-------|-------|-------|------|------|------|------|------|------|------|------|-------|----|
| A | 0.22 | 0.57 | 0.16 | 0.12 | 0.10 | 0.54 | 0.28 | 0.40 | 0.37 | 0.53 | 0.53 | 0.35 | 0.24 | 0.76 | 0 | 0 | 0 |
| B | 6.14 | 6.25 | 3.57 | 2.18 | 1.69 | 6.80 | 3.65 | 3.30 | 6.27 | 1.43 | 6.77 | 3.55 | 2.03 | 5.15 | 2.60 | 2.40 | 0 |
| C | 0.08 | 0.23 | 0.06 | 0.06 | 0.44 | 0.77 | 0.17 | 0.04 | 0.09 | 0.05 | 0.13 | 0.09 | 0.03 | 0.18 | 0 | 0 | 0 |
| D | 1.48 | 2.27 | 1.45 | 0.37 | 4.01 | 12.67 | 5.83 | 1.98 | 0.58 | 0.45 | 3.30 | 1.54 | 0.82 | 3.53 | 0 | 0 | 0 |
| E | 0 | 0 | 0 | 0.03 | 0 | 0.10 | 0.14 | 0 | 0.03 | 0 | 0.17 | 0 | 0.01 | 0.07 | 0 | 0 | 0 |
| F | 0 | 0.23 | 0.03 | 0.01 | 0.07 | 0.23 | 0.18 | 0.09 | 0.03 | 0.01 | 0.07 | 0.07 | 0.03 | 0.04 | 0.20 | 0 | 0 |
| G | 0.14 | 0.23 | 0.93 | 0.15 | 0.20 | 4.29 | 1.44 | 0.13 | 0.55 | 0.45 | 0.50 | 0.29 | 0.24 | 2.81 | 0.40 | 0 | 0 |
| H | 8.53 | 8.64 | 72.59 | 3.80 | 2.87 | 17.43 | 52.45 | 2.47 | 4.21 | 2.64 | 6.84 | 4.04 | 2.08 | 7.06 | 0.60 | 17.60 | 0 |
| J | 0.14 | 0.23 | 0.93 | 0.18 | 2.78 | 1.41 | 11.97 | 0.09 | 0.28 | 0.69 | 0.17 | 0.07 | 0.18 | 0.47 | 0 | 0 | 0 |
| K | 0.05 | 0.23 | 1.48 | 0.25 | 81.72 | 1.26 | 6.35 | 0.04 | 0.34 | 0.27 | 0.23 | 0.53 | 0.24 | 0.32 | 0 | 0 | 0 |
| L | 0.11 | 0.23 | 0.74 | 0.23 | 0.43 | 1.59 | 0.71 | 0.40 | 0.46 | 0.21 | 0.53 | 0.29 | 0.26 | 3.03 | 1.00 | 0 | 0 |

| IDs | 1 | 2 | 3 | 4 | 5 | 6 | 7 | 8 | 9 | 10 | 11 | 12 | 13 | 14 | 15 | 16 | 17 |
|---|---|---|---|---|---|---|---|---|---|---|---|---|---|---|---|---|---|
| M | 0.03 | 0.11 | 0.06 | 0.10 | 0.02 | 1.08 | 0.15 | 0.04 | 0.18 | 0 | 0.10 | 0.44 | 0.07 | 0 | 0 | 0 | 0 |
| N | 0.03 | 0.23 | 0.29 | 0.18 | 0.05 | 1.62 | 0.09 | 0.09 | 0.43 | 0.10 | 0.10 | 0.13 | 0.20 | 0.79 | 0 | 0 | 0 |
| P | 3.95 | 6.25 | 4.31 | 3.05 | 3.60 | 43.57 | 7.24 | 7.62 | 6.79 | 2.70 | 6.21 | 4.52 | 3.38 | 8.18 | 6.60 | 24.00 | 0 |
| Q | 36.18 | 64.77 | 7.40 | 34.36 | 0.57 | 2.25 | 4.78 | 71.33 | 66.44 | 36.53 | 52.42 | 67.24 | 84.37 | 49.32 | 74.80 | 16.80 | 100 |
| R | 1.07 | 8.07 | 0.87 | 0.49 | 0.38 | 0.84 | 0.78 | 10.61 | 9.22 | 0.64 | 19.35 | 15.80 | 1.33 | 14.34 | 4.60 | 9.60 | 0 |
| S | 3.48 | 0.11 | 3.25 | 0.37 | 0.16 | 1.06 | 0.92 | 0.44 | 0.09 | 50.23 | 0.60 | 0.42 | 0.96 | 1.19 | 9.20 | 2.40 | 0 |
| T | 38.37 | 1.36 | 1.83 | 54.06 | 0.82 | 1.95 | 2.79 | 0.92 | 3.63 | 3.05 | 1.70 | 0.64 | 3.51 | 2.63 | 0 | 27.20 | 0 |
| U | 0 | 0 | 0.03 | 0.02 | 0.07 | 0.16 | 0.04 | 0 | 0 | 0 | 0 | 0 | 0.00 | 0.07 | 0 | 0 | 0 |
| V | 0 | 0 | 0 | 0 | 0 | 0 | 0 | 0 | 0 | 0 | 0 | 0 | 0.00 | 0 | 0 | 0 | 0 |
| Z | 0 | 0 | 0 | 0 | 0.03 | 0.38 | 0.05 | 0 | 0 | 0.01 | 0.27 | 0.02 | 0.03 | 0.07 | 0 | 0 | 0 |

Table 7

*ID-Faculty, ID-Category, and ID Influencer*

| ID | Faculty Name | ID | Category | ID Influencer |
|---|---|---|---|---|
| 1 | Faculty of Agro-Industry | A | General Works | 10,2,8,14,12 |
| 2 | Faculty of Dentistry | B | Philosophy, Psychology, Religion | 6,2,11,14,9,8,7 |
| 3 | Faculty of Economics | C | Auxiliary Science of History | 6 |
| 4 | Faculty of Engineering | D | World History | 6,7,5,14,11 |
| 5 | Faculty of Law | E | History of the Americas | - |
| 6 | Faculty of Liberal Arts | F | History of the Americas | - |
| 7 | Faculty of Management Sciences | G | Geography, Anthropology, Recreation | 6,14,7 |
| 8 | Faculty of Medical Technology | H | Social Sciences | 3,7,6 |
| 9 | Faculty of Medicine | J | Political Sciences | 7,5 |
| 10 | Faculty of Natural Resources | K | Law | 5,7 |
| 11 | Faculty of Nursing | L | Education | 14,6,3 |
| 12 | Faculty of Pharmaceutical Sciences | M | Music and Books on Music | - |
| 13 | Faculty of Science | N | Fine Arts | 6,14 |
| 14 | Faculty of Traditional Thai Medicine | P | Language and Literature | 6,16,14,8,7,9 |
| 15 | Faculty of Veterinary Science | Q | Science | 13,12,15,8,2,9,11 |
|  |  | R | Medicine | 11,16,14,9,8,12,2 |
|  |  | S | Agriculture | 10,15,1 |
|  |  | T | Technology | 4,1 |
|  |  | U | Military Science | - |
|  |  | V | Naval Science | - |
|  |  | Z | Bibliography, Library Science, etc. | - |

## Analysis of patron's grade level vs. books categories

The focus of this part is *"Is there a relationship between library collections usage and achievement of student within a given faculty?"*. Analysis of the patron's book loan and patron's cumulative grade point average (CGPA) during academic years 2015-2018 were carried out to study how patron's academic performance influences their book category choices. A total of 64,053 transactions of circulation data and 6,534 records of student record data who borrowed books were collected. The student achievement data (Cumulative GPA) was divided into 5 groups based on grade level, i.e. *Excellent (3.50-4.00), Very Good (3.00-3.49), Good (2.50-*

*2.99), Average (2.00-2.49), and Poor (< 2.00).* Figure 2 shows the percentage of items check-out distributed by patron's grade level and LC category. The results revealed that most books borrowed by patrons in *"Good"* and *"Very Good"* grade level fall in category *Q: Science*. It is also interesting that the number of items check-out from patron's grade level in *"Good"* and *"Very Good"* are between 30 and 50 items (Figure 3). This is similar to the result of Renaud et al. (2015) that a large number of students with higher GPAs is correlated with higher library check-out activity.

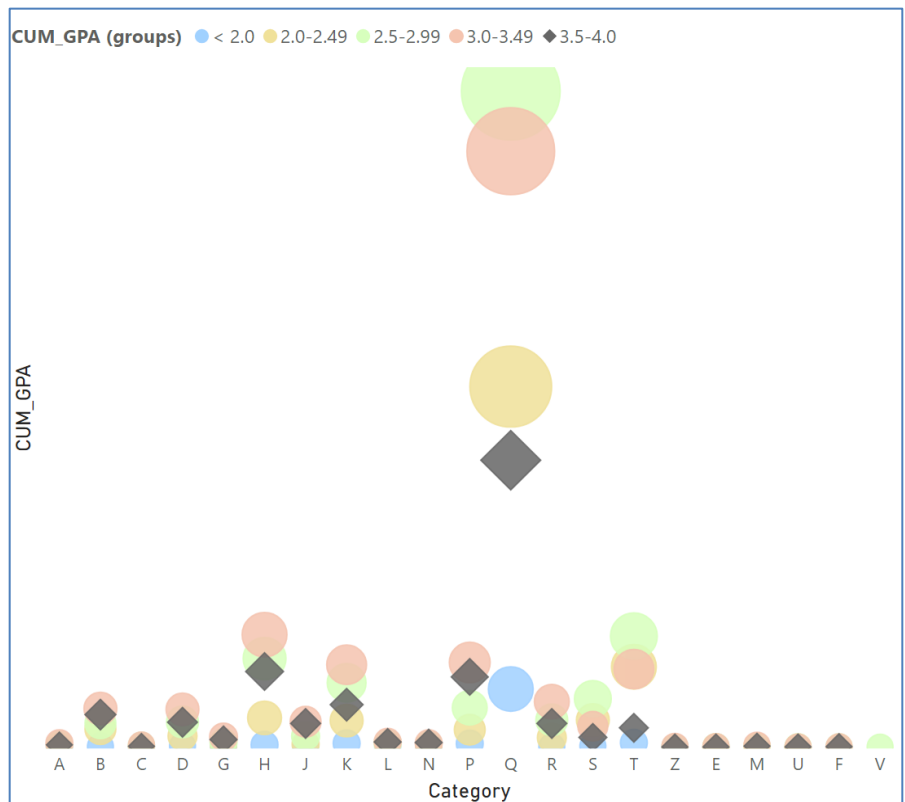

*Figure 2:* Percentage of items check-out distributed by patron's grade level and LC category

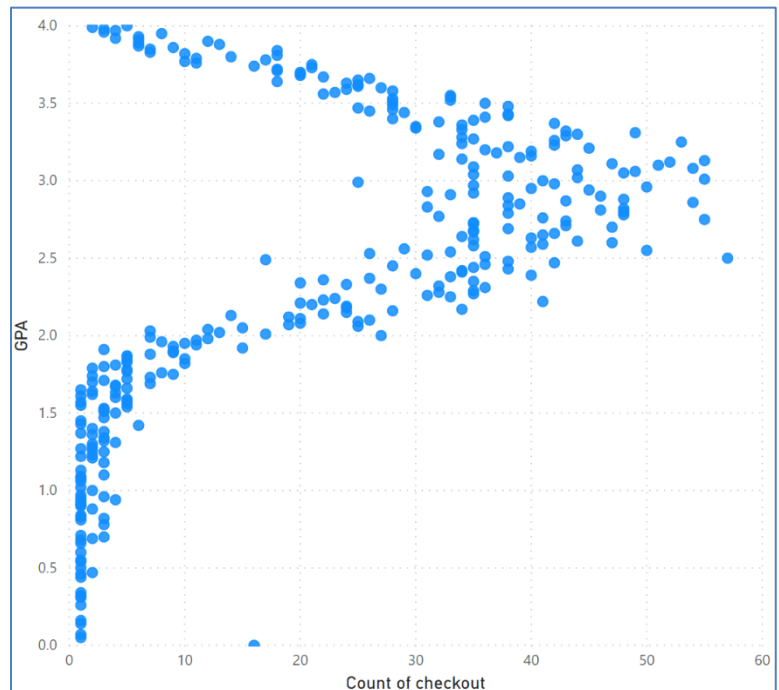

*Figure 3:* Scatter plot between number of items check-out and patron's CGPA

**Analyses of book usage**

Circulation data can help libraries determine which book category to keep on site or withdraw from the collection. The objective of this part is to find the useful lifespan of books and how long we should relocate them to isolated storage. First, book usage of each category and book's useful lifespan were calculated from book data. Figure 4presentsthe distribution of the books, based on LC category, relationship between theirs loans and not-loans. The results show that the percentage of borrowing books is close to the percentage of uncirculated books. It is interesting to observe that book usage patterns in terms of number and subject area, books in the LC category *B (philosophy, psychology, religion)*, category *G (geography, anthropology, recreation)*, category *H (social sciences)*, category *K (law)*, category *M (music and books on music)*, category *P (language and literature)*, category *Q (science)*, and the *category V (naval science)* have significant number of check outs.

To determine the useful lifespan of a book, the duration between the date of cataloging to the date of last check-in is considered. A book's useful lifespan is calculated from the total number of days between the cataloging date and the date of last check-in. Figure 5 reveals how the useful life of books varies on the LC classification. It is found that books related to language and literature (category *P*) have longer useful lifespan than books in the sciences (category *Q*) and technology (category *T*). It is also evident that books related to the social sciences (category *H*), philosophy (category *B*), music (category *M*), and fine arts (category *N*) are all used less time than the sciences (category *Q*). There are many books in some collection that have never been checked out, these uncirculated items are removed from our investigation. Line chart of Figure 6shows the results of removed items that are never circulated. The LC category *Q: Science*, category *S: Agriculture*, and category *P: Language and Literature* rise significantly. It is assumed that as circulation data grow over the past 20 years, a difference in book use will be revealed between the sciences and the humanities. Books in the LC category *Q: Science* take circulation over the longest period. Books in category *P: Language and Literature* are closely trailed by books in the category *Q: Science*. Books in the LC category *U: Military Science* show the greatest number of days of circulation after cataloging among other classes, while; books in the LC category *V: Naval Science* show the least number of days of circulation after cataloging.

Lastly, the widely influential patterns were examined using the useful lifespan of each item category. Figure 6 shows the percentage of books in each category having a lifespan of 0-10 year, and 11-20 years after removing the uncirculated items. It is clear from Figure 6 that the patterns of useful books have changed significantly over the past 20 years. The books in LC category *Q: Science* and category *P: Language and Literature* show the longer lifespan than books in other classes. While having of 10 years books lifespan, items usefulness falls into category *V: Naval Science*, category *K: Law*, and category *B: Philosophy, Psychology, and Religion,* respectively.

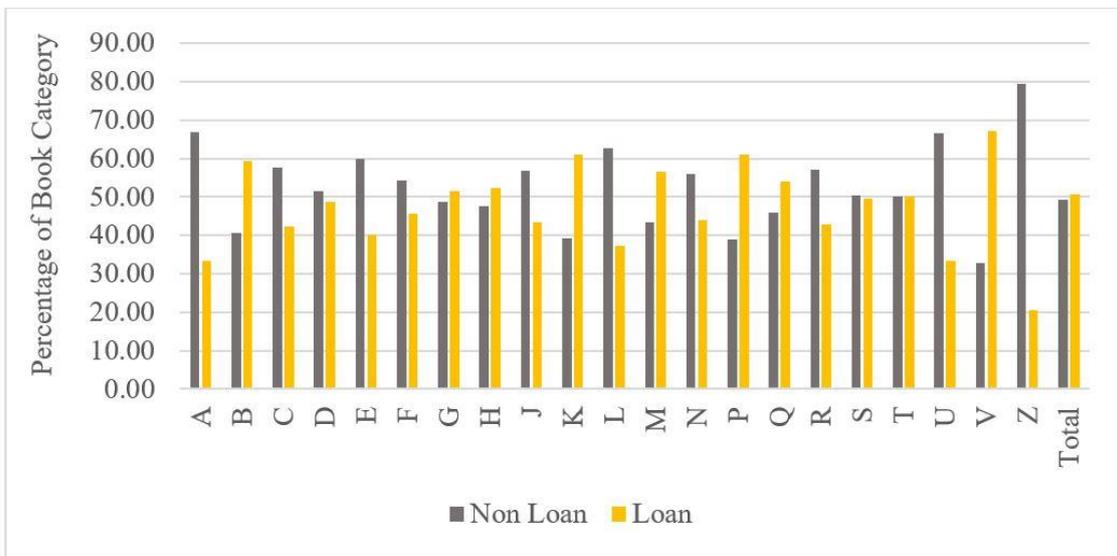

*Figure 4:* Check-out distribution by LC category

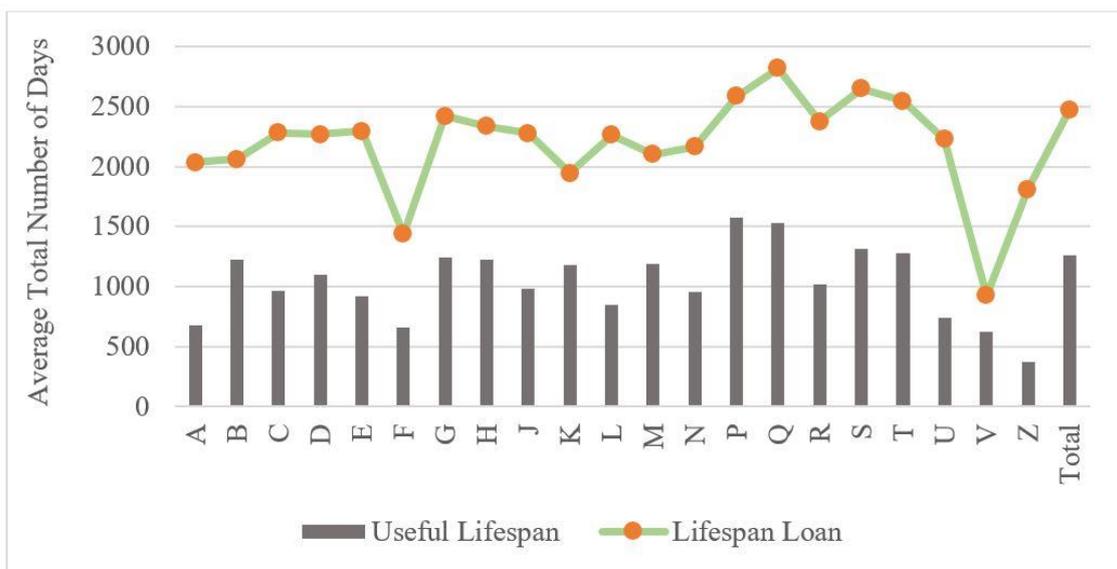

*Figure 5:* LC category lifespan distribution

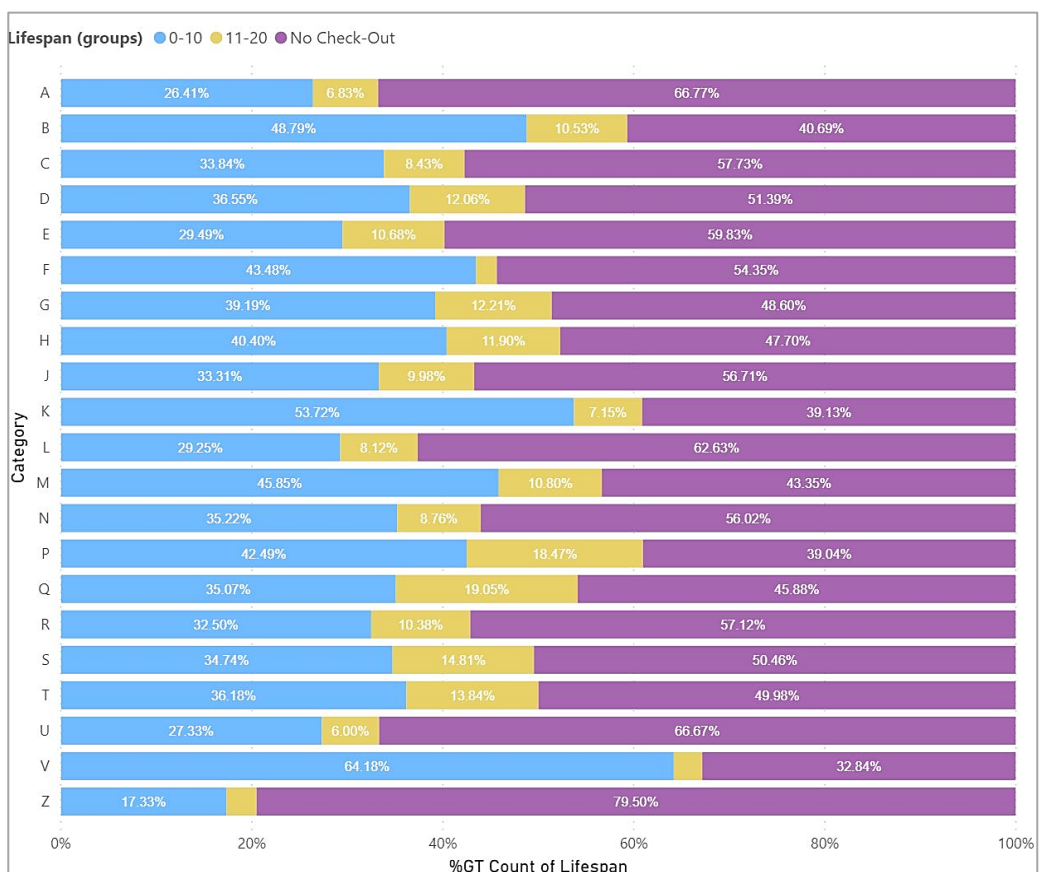

*Figure 6:* Percentage LC category lifespan distribution

**Predictive Data Mining Model**

In this part, the questions are "*What are the association rules when patrons borrow books?*" and "*Is there relationship between student academic achievement and theirs book loans?*". In order to answer these two questions, the expectations for this investigation and all the details of information are summarized in Table 8. This study collects 6,534records of a circulation databetween August 2015 and July 2018 from ALIST system and a student record data duringacademic years 2015-2018 from PSU Student Information System. WEKA software, a data mining tool, was used to analyze the data to discover association rules and clustering.

Table 8
*Description of the problems by data mining task*

| Problem Solving | Expectation | Data Source | Model | Tool |
|---|---|---|---|---|
| *1. What are the association rules when patrons borrow books?* | Understand the association rules of different faculty students may have different needs and behavior patterns. | 6,534 records of undergraduate students borrowed books between August 2015 and July 2018 | Association Rules | WEKA 3.8.3 |
| *2. Is there relationship between student academic achievement and theirs book loans?* | Understand the relationship between student academic achievement and their required books. | Data during academic years 2015-2018 are from (1) 64,053 records of undergraduate students'circulation data in 17 faculties (2) 104,066records of all undergraduate transactions | Cluster | |

*Analyses patron interesting by Apriori*

The goal of this part is to generate association rules in a circulation data set. This work uses the *Apriori* algorithm as a methodology to find the association rules of a data source given in Table 9.A set of initial *Apriori* parameters were selected and input to WEKA tool, for example, minimum support, minimum confidence, and number of rules as 0.1, 0.9, 10 respectively. The experimental setup selected for mining rules using *Apriori* is given in Table 9.The count of best rules found based on the correlation among attributes for the instances in the experiment conducted (Table 10).Figure 7shows the association rules of undergraduate students who borrowed books during academic years 2015-2018. In the list of best rules found, for example, the first rule (R1) means if a student borrows the category books of *QA: Mathematics*, *QC: Physics*, *QH: Biology*, and *QR: Microbiology* then that student may also borrow the books of *QD: Chemistry*. The rule R3 means if a Faculty of Science student borrows books of category *QC: Physics*, *QH: Biology*, and *QR: Microbiology* then this science student may also borrow the books of QD: *Chemistry*. One interesting finding is that if a student borrows the books of *TA: Civil Engineering* and *TK: Electrical Engineering* (R9), *TA: Civil Engineering* and *TJ: Mechanical Engineering* (R11), or *QA: Mathematics* and *TJ: Mechanical Engineering* (R18) then that student will be an Engineering student. The other interesting rules inferred in R17 provide information of the books selected by students from Faculty of Pharmaceutical Sciences like books of category *QA: Mathematics*, then those students will also borrow books of category *QD: Chemistry*.

The interesting rules given in Figure 8shows that books of category *QM: Human Anatomy* and category *QP: Physiology*; category *QD: Chemistry* and category *QM: Human Anatomy*; and category *PE: English* are likely borrowed by students from Faculty of Dentistry. Another interesting rule pattern found in Figure 9 is the borrowing pattern of Faculty of Economics students which shows preferential on books of category *HB: Economic theory, Demography* and category *HF: Commerce*; category *HG: Finance*; category *HD: Industries, Land Use, Labor*; and category *HB: Economic theory, Demography* and category *QA: Mathematics*.

Table 9
*Apriori parameters setup*

| Data set | Instances | Attributes | Minimum Support | Minimum Confidence | Best rules found |
|---|---|---|---|---|---|
| Patron's Book-Loan | 6534 | 188 | 0.01 | 0.9 | 25 |
| Faculty of Agro-Industry | 183 | 188 | 0.1 | 0.9 | 15 |
| Faculty of Dentistry | 75 | 188 | 0.1 | 0.9 | 8 |
| Faculty of Economics | 170 | 188 | 0.1 | 0.9 | 7 |
| Faculty of Engineering | 970 | 188 | 0.1 | 0.9 | 14 |
| Faculty of Law | 293 | 188 | 0.1 | 0.9 | 5 |
| Faculty of Liberal Arts | 387 | 188 | 0.1 | 0.9 | 7 |
| Faculty of Management Sciences | 741 | 188 | 0.1 | 0.9 | 6 |
| Faculty of Medical Technology | 139 | 188 | 0.3 | 0.9 | 5 |
| Faculty of Medicine | 390 | 188 | 0.1 | 0.9 | 13 |
| Faculty of Natural Resources | 442 | 188 | 0.1 | 0.9 | 10 |
| Faculty of Nursing | 409 | 188 | 0.1 | 0.9 | 10 |
| Faculty      of      Pharmaceutical | 351 | 188 | 0.3 | 0.9 | 3 |

| Data set | Instances | Attributes | Minimum Support | Minimum Confidence | Best rules found |
|---|---|---|---|---|---|
| Sciences | | | | | |
| Faculty of Science | 1641 | 188 | 0.1 | 0.9 | 19 |
| Faculty of Traditional Thai Medicine | 276 | 188 | 0.1 | 0.9 | 15 |
| Faculty of Veterinary Sciences | 49 | 188 | 0.3 | 0.9 | 6 |
| International College | 16 | 188 | 0.1 | 0.9 | 6 |
| Faculty of SINO-Thai | 2 | 188 | 0.1 | 0.9 | 2 |

```
=== Run information ===
Scheme:      WEKA.associations.Apriori -N 50 -T 0 -C 0.9 -D 0.05 -U 1.0 -M 0.01 -S -1.0 -c -1
Relation:    asso_subclass-WEKA
Instances:   6534
Attributes:  188
        [list of attributes omitted]
=== Associator model (full training set) ===
Apriori
=======

Minimum support: 0.01 (65 instances)
Minimum metric <confidence>: 0.9
Number of cycles performed: 20

Generated sets of large itemsets:
Size of set of large itemsets L(1): 58
Size of set of large itemsets L(2): 124
Size of set of large itemsets L(3): 83
Size of set of large itemsets L(4): 24
Size of set of large itemsets L(5): 3

Best rules found:

 1. QA=t QC=t QH=t QR=t 67 ==> QD=t 66   <conf:(0.99)> lift:(2.67) lev:(0.01) [41] conv:(21.15)
 2. QA=t QC=t QD=t QR=t 68 ==> QH=t 66   <conf:(0.97)> lift:(3.39) lev:(0.01) [46] conv:(16.18)
 3. Sci=t QC=t QH=t QR=t 87 ==> QD=t 84   <conf:(0.97)> lift:(2.62) lev:(0.01) [51] conv:(13.73)
 4. QA=t QH=t QR=t 102 ==> QD=t 98   <conf:(0.96)> lift:(2.61) lev:(0.01) [60] conv:(12.88)
 5. Sci=t QC=t QR=t 96 ==> QD=t 92   <conf:(0.96)> lift:(2.6) lev:(0.01) [56] conv:(12.12)
 6. QA=t QC=t QR=t 71 ==> QD=t 68   <conf:(0.96)> lift:(2.6) lev:(0.01) [41] conv:(11.21)
 7. Sci=t QA=t QR=t 69 ==> QD=t 66   <conf:(0.96)> lift:(2.59) lev:(0.01) [40] conv:(10.89)
 8. Sci=t QA=t QR=t 69 ==> QH=t 66   <conf:(0.96)> lift:(3.34) lev:(0.01) [46] conv:(12.31)
 9. TA=t TK=t 68 ==> Eng=t 65   <conf:(0.96)> lift:(6.44) lev:(0.01) [54] conv:(14.48)
10. QA=t QC=t QR=t 71 ==> QH=t 67   <conf:(0.94)> lift:(3.3) lev:(0.01) [46] conv:(10.14)
11. TA=t TJ=t 88 ==> Eng=t 83   <conf:(0.94)> lift:(6.35) lev:(0.01) [69] conv:(12.49)
12. QA=t QR=t 124 ==> QD=t 116   <conf:(0.94)> lift:(2.54) lev:(0.01) [70] conv:(8.7)
13. QA=t QC=t QR=t 71 ==> QD=t QH=t 66   <conf:(0.93)> lift:(5.07) lev:(0.01) [52] conv:(9.66)
14. QC=t QH=t QR=t 126 ==> QD=t 117   <conf:(0.93)> lift:(2.52) lev:(0.01) [70] conv:(7.95)
15. QC=t QL=t 83 ==> QH=t 77   <conf:(0.93)> lift:(3.24) lev:(0.01) [53] conv:(8.46)
16. QC=t QD=t QL=t 75 ==> QH=t 69   <conf:(0.92)> lift:(3.21) lev:(0.01) [47] conv:(7.65)
17. Pharma=t QA=t 85 ==> QD=t 78   <conf:(0.92)> lift:(2.49) lev:(0.01) [46] conv:(6.71)
18. QA=t TJ=t 72 ==> Eng=t 66   <conf:(0.92)> lift:(6.17) lev:(0.01) [55] conv:(8.76)
19. Sci=t QC=t QD=t QR=t 92 ==> QH=t 84   <conf:(0.91)> lift:(3.19) lev:(0.01) [57] conv:(7.3)
20. Sci=t QA=t QC=t QH=t 241 ==> QD=t 220   <conf:(0.91)> lift:(2.48) lev:(0.02) [131] conv:(6.92)
21. Sci=t QC=t QR=t 96 ==> QH=t 87   <conf:(0.91)> lift:(3.17) lev:(0.01) [59] conv:(6.85)
22. QL=t QR=t 74 ==> QH=t 67   <conf:(0.91)> lift:(3.16) lev:(0.01) [45] conv:(6.6)
23. QC=t QL=t 83 ==> QD=t 75   <conf:(0.9)> lift:(2.45) lev:(0.01) [44] conv:(5.82)
24. Sci=t QC=t QL=t 112 ==> QH=t 101   <conf:(0.9)> lift:(3.15) lev:(0.01) [68] conv:(6.66)
25. QA=t QC=t QH=t 284 ==> QD=t 256   <conf:(0.9)> lift:(2.44) lev:(0.02) [151] conv:(6.18)
```

*Figure 7*: Association rules of patrons borrowed books during academic years 2015-2018

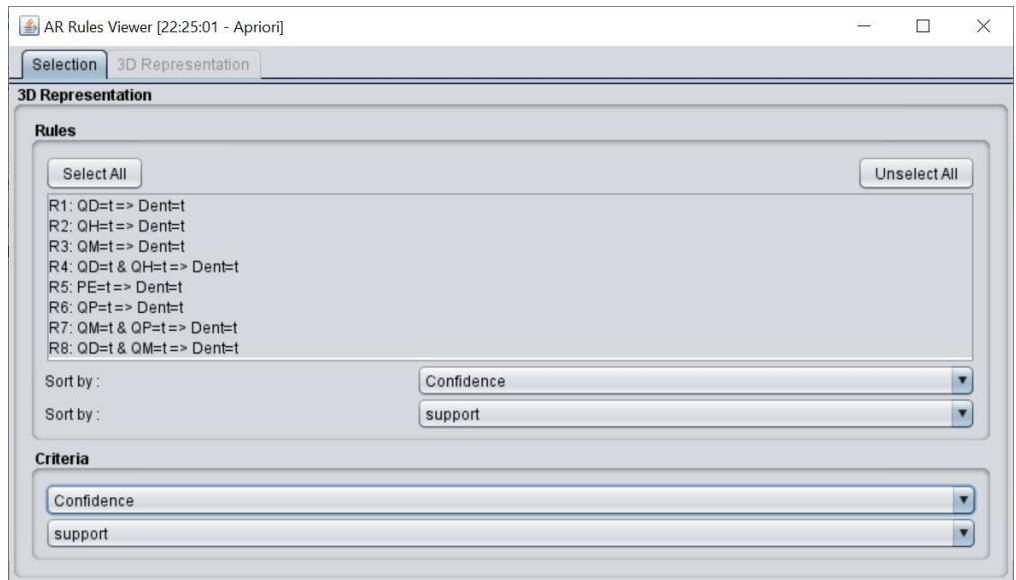

*Figure 8.* Rules pattern found of Faculty of Dentistry

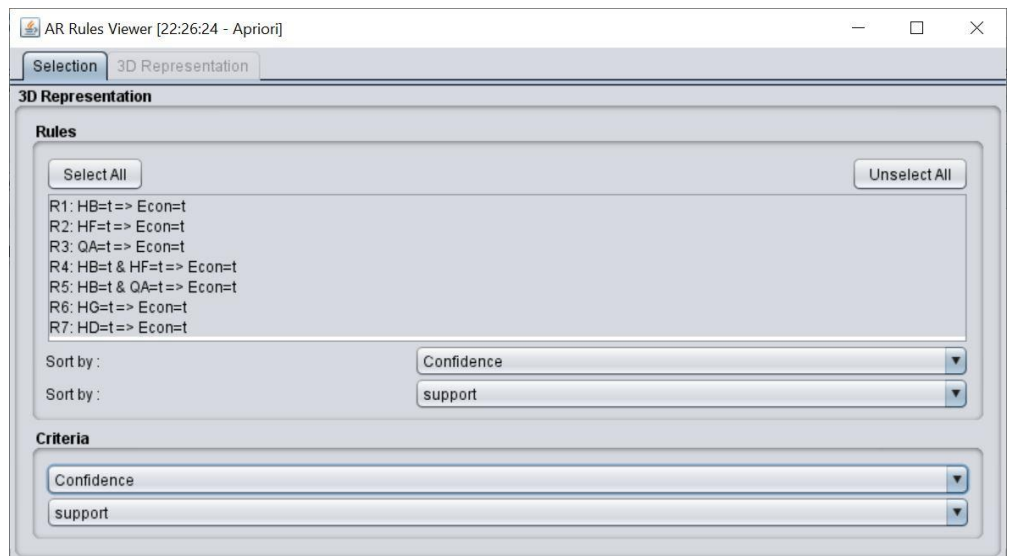

*Figure 9.* Rules pattern found of Faculty of Economics

### *Analyses of patron characteristics by K-Means*

Another research question relates to library usage and student academic achievement. Clustering analysis was done to identify relations of patron book loans and undergraduate students' CGPA.

*1) Clustering of patron's faculties*

Table 10 shows that the most demanding books of Science students are those of category *QD: Chemistry* and mostly with lifespan of 16-20 years (cluster $c_1$) or of the category *QH: Biology* with 6-10 years lifespan (cluster $c_3$). Patrons in cluster $c_2$ are Law students who prefer books of category *KP: Law in Asia & Eurasia, Africa, and Pacific Asia & Antarctica* and demanding a lifespan of 6-10 years. Patron's behavior in cluster $c_4$ are Engineering students mostly use books of category *QA: Mathematics* and lifespan of 11 to 15 years. Cluster $c_5$ includes Management Sciences students who prefer books of category *HB: Economic Theory, Demography* with the shortest lifespan (0-5 years). This is similar to the results of Wang et al.

(2011) that patrons in different departments may have different interests in borrowing books. The School of Business Administration has more frequent borrowing than other departments.

Table 10
*Characteristics of patron's book loans based on book lifespan and subcategories*

| Attribute | Full Data | $c_1$ | $c_2$ | $c_3$ | $c_4$ | $c_5$ |
|---|---|---|---|---|---|---|
| Faculty | Faculty of Science | Faculty of Science | Faculty of Law | Faculty of Science | Faculty of Engineering | Faculty of Management Sciences |
| Subcategory | QD | QD | KP | QH | QA | HB |
| Book Lifespan | 11-15 | 16-20 | 0-5 | 6-10 | 11-15 | 0-5 |

### 2) Clustering of patron's grade levels

The results show that there are relationships between student academic achievements and the category of books loaned. The relations among faculty, subcategory and grade level were obtained by using *K*-Means algorithm and shown in Table 11. It is evident from the results that patrons from Faculty of Science borrowed books of category *QD: Chemistry* and had CGPA 3.00-3.49 (cluster $c_1$).Another group of Sciences students borrowed books of category *QH: Biology* and had CGPA 2.50-2.99 (cluster $c_3$). Cluster $c_2$ is another group of patrons who are Law students with grade point average 2.00-2.49 demanding books of category *KP: Law in Asia & Eurasia, Africa, Pacific, Asia & Antarctica*. Patron's behavior in clusters$c_4$and $c_5$are similar in academic achievement, but different in the category of books, i.e. patrons in cluster $c_4$ (Management Sciences students with grade point average 3.00-3.49) prefer books of *category PE: English*, while patrons in cluster $c_5$favor books of category *HB: Economic Theory, Demography*.

Regarding patron's book loans, besides the clustering of patron's academic achievements, the result of clustering to patron's borrowing frequency of all grade levels was shown as Table 12. The result shows that books of category *HB: Economic Theory, Demography* fall into the group of excellent students most frequently used books (cluster $c_5$). The English books (category *PE: English*) fall into the group of poor students borrowed book category (cluster $c_4$).This is similar to the results of Wang et al. (2011) that the English books fall into the heavily used book. This could be reading foreign books is an important way to learn a foreign language and improve the English level. Moreover, English books are popular among lower-grade students because the PSU exit exam's policy requires an English test for all students to complete their degree.

Table 11
*Characteristics of the patron's book loans based on grade levels and subcategories*

| Attribute | Full Data | $c_1$ | $c_2$ | $c_3$ | $c_4$ | $c_5$ |
|---|---|---|---|---|---|---|
| Faculty | Faculty of Science | Faculty of Science | Faculty of Law | Faculty of Science | Faculty of Management Sciences | Faculty of Management Sciences |
| Subcategory | QD | QD | KP | QH | PE | HB |
| Grade Level | Good | Very Good | Average | Good | Very Good | Very Good |

Table 12
*Characteristics of the patron's book loans in grade levels classes to clusters*

| Attribute | $c_1$ | $c_2$ | $c_3$ | $c_4$ | $c_5$ |
|---|---|---|---|---|---|
| Subcategory | QD | KP | QH | PE | HB |
| Grade Level | Good | Very Good | Average | Poor | Excellent |

## Conclusions and Future Work

This study used data mining to analyze the real circulation detain Khunying Long Athakravisunthorn Learning Resources Center. The *Apriori* and *K*-Means algorithm were used to mine the association rules and clustering of the library datasets. 64,053 records between August 2015 and July 2018 were collected from two data sources: academic library system and student information system. A data mining tool, WEKA, was also used to find the association rules and clustering of transaction data sets. Data mining is an important need for libraries to bring out the hidden knowledge.

The results of association rule mining in patron's book-loan data reveal interesting rules for the patron's interest in the usage of books. One interesting finding is that different faculty students have different interesting and behavior. For example, if a dentistry student borrowed a book of the second level categories *QD, QH, QM, QP*, and then that student also borrowed subcategory *PE*. Economics students who borrowed books of categories *HB, HF, HG, HD*, then they also borrowed books belong to Mathematics (*QA*).A significant result suggests that the library can establish the associated bookshelf for patron's selection by their association. The circulation patterns among the different disciplines remains an area for further research. The next level of analysis may review check-out patterns more narrowly divided by departments, majors, and curriculum. To determine that association rules of different departments, majors, and curriculum students may have different needs and behavior patterns from the initial findings. The rules also help the library for arriving decision related to investment of budget for specific books titles. Other academic libraries may benefit from this study to improve user services based on data patterns. Conducting this research at several universities may reveal similarities that can be generalized to other academic libraries. Moreover, this study has carried out the analysis of association rules only for books which can be further extended for other materials. In technical aspect, the multiple-level association rules will provide better understanding of patron usage patterns for taking an effective decision which will be considered further research.

The results of the cluster analysis reflect the relationship between categories of books borrowed and their cumulative grade point average (CGPA). For example, patrons from Faculty of Management Sciences and having grade point average in range of 3.00-3.49 prefer to borrow books of subcategory *PE* and *HB*. The library can recommend books to the patrons according to their library usage and achievements. It can be said that these achieved clusters can encourage librarians to carry out further studies by understanding the relationship between student academic achievement and their required books. Books also are needed in some disciplines more than others. Breakdown of the book-loan data into specific majors/curriculum may reveal a proper relationship between library use and student achievement. Many other possible factors in library support services and activities, for example, advising services, user education, and use of tutoring may contribute to student achievement. Therefore, possible factors may be analyzed to determine how those relationships compare to student achievement and library usage.

From the experimental results, it can be found that data mining techniques helps to identify the patterns, by discovering the hidden knowledge stored in the databases for using the libraries more effectively and efficiently. The various techniques of data mining like association rules and clustering can be applied to bring out hidden knowledge from the libraries and information services. Many foreign academic libraries have been successfully applying the data mining technique to work and management. These libraries may not only be the collection of traditional institutions but also become information institutions. Also, these library staff no longer are in the traditional significance librarian, they may develop into the information expert.

Considering the circulation analysis on book borrowing, the library can consider about how to arrange book purchasing, shelving and maintenance and also consider about how to use data mining approach for analysis of the book collections, development of patrons need, and interest of library services offered. The increase in library use will lead to a change of the library marketing to attract the patrons and will enable the new creation of standardsin library services.